\documentclass{aastex}
\usepackage{spr-astr-addons}
\usepackage{url}

\RequirePackage{color}

\newcommand{\emaila}{wilhelm@mps.mpg.de}
\newcommand{\emailb}{bholadwivedi@gmail.com}

\newcommand*\Del{\mathrm{\Delta}}                 

\newcounter{saveeqn}

\begin{document}

\title{The Aharonov--Bohm effect: A quantum or a relativistic phenomenon?}

\shorttitle{The Aharonov--Bohm effect}
\shortauthors{K. Wilhelm and B.N. Dwivedi}

\author{Klaus Wilhelm}
\affil{Max-Planck-Institut f\"ur Son\-nen\-sy\-stem\-for\-schung
(MPS), 37077 G\"ottingen, Germany\\ \emaila}
\and
\author{Bhola N. Dwivedi}
\affil{Department of Physics, Indian Institute of Technology
(Banaras Hindu University), Varanasi-221005, India \\ \emailb}

\today

\vspace{1cm}

\begin{abstract}
The Aharonov--Bohm effect is considered by most authors as a quantum effect,
but a generally accepted explanation does not seem to be available. The
phenomenon is studied here under the assumption that hypothetical electric
dipole distributions configured by moving charges in the solenoid act on the
electrons as test particles. The relative motions of the interacting charged
particles introduce relativistic time dilations. The massless dipoles are
postulated as part of an impact model that has recently been proposed to
account for the far-reaching electrostatic forces between charged particles
described by Coulomb's law. The model provides a quantitative explanation of
the Aharonov--Bohm effect.
\end{abstract}


\section{Introduction}
\label{introd}

The Aharonov--Bohm effect was theoretically predicted in the middle of the
last century \citep{EhrSid,AhaBoh}. It was, however, implicitly derived by
\citet{Gla33} from Fermat's principle and the refractive index of electron
optics in 1932. Many experimental verifications have been performed since
1960 \citep[e.g.][]{Cha60,MoeBay,Len62}; see also \citet{Tonetal} and
\citet{Capetal} for further references. Most authors consider the effect to
be a purely quantum mechanical one, but there are opposing views
\citep[e.g.][]{Boy08}. This has been summarized in a recent paper by
\citet{HegNeu}, stating that a generally accepted physical understanding of
the Aharonov--Bohm effect does not seem to be available. Here we will outline
a solution that is based on a relativistic effect.
%
\begin{figure}[t]
\vspace*{2mm}
\begin{center}
\includegraphics[width=8.3cm]{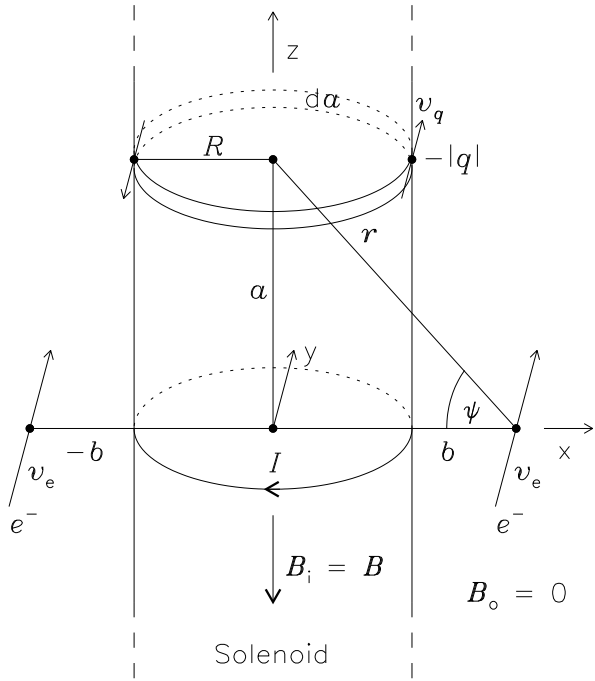}
\end{center}
\caption{Schematic diagram of the Aharonov--Bohm effect.
A very extended solenoid along the $z$ axis,
with radius $R$ and $N = n\,a$ windings on an axial length, $a$ (i.e.,
a winding density of $n$),
carrying a current, $I$, (realized by negative charges, $-\,|q|$, moving with a
speed $v_q$) has an internal magnetic flux density of
$B_{\rm i} = \mu_0\,n\,I$ and nearly no magnetic field, $B_{\rm o}$, outside.
Nevertheless, electrons travelling with the velocity $v_{\rm e}$ parallel to
the y axis on either side of the coil at impact parameters $x = \pm\,|b|$
experience a shift of their interference pattern
on a screen behind the plane of the paper (cf., Fig.~\ref{solenoid}).
\label{configuration}}
\end{figure}

\section{Outline of a solution} 
\label{aha}

The Aharonov--Bohm effect, as illustrated and described in
Fig.~\ref{configuration}, can be understood in the context of an electric
dipole model proposed by \citet{Wiletal} for electrostatic forces. According
to this model and its application to magnetostatic configurations
\citep{Dwietal}, the dipole distributions would differently be modified
inside and outside of the solenoid. \emph{Of particular importance is that the
outside will indeed affected at all, in contrast to the zero magnetic field
there under ideal conditions}. Charged particles (electrons are generally
used as test particles) moving outside of the solenoid in a plane
perpendicular to its axis will react to the modified dipole distributions.
On one side of the coil, a component of the velocity, $v_q$, of the charge
carriers in the conductor will be in the same direction as the velocity
vectors of the electrons, $\mbox{\boldmath$v$}_{\rm e}$, and on the other side
it is oppositely directed. This is shown in Fig.~\ref{cross_section} and will
lead to different relativistic time dilations. We thus expect slightly
different integrated momentum transfers in the plane of motion normal to both
the axis of the coil and the trajectories of the electrons.
Note that no significant acceleration or deceleration would occur along the
trajectory of the test particle in line with observations. However, the
transverse momentum transfers are at variance with a dispersionless
interaction of the Aharonov--Bohm effect and a propagation of the average
positions of the electron wave packets as if they would travel in free
space \citep{Capetal}, cf., also \citet{McGetal} and \citet{McG13}.

\section{Quantitative description of the effect} 
\label{Quantitative}

In order to arrive at a quantitative description of the effect, several
simplifying assumptions and approximations (considered to be reasonable) have
to be made. This limits the calculations to an amount acceptable within this
short communication. A number of $Z$ negative charges, $- |q|$, is assumed to
be distributed along the circumference of one winding of the solenoid. The
moving negative charges (i.e., electrons) constitute a current
%
\begin{figure}[t]
\vspace*{2mm}
\begin{center}
\includegraphics[width=8.3cm]{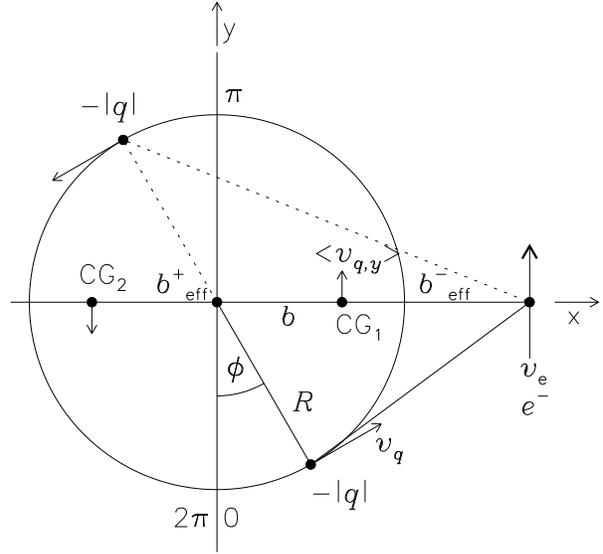}
\end{center}
\caption{Cross-section of the solenoid in the $(x,y)$ plane. An electron
passing with a velocity $v_{\rm e}$ at $x = b$ is indicated. The current in
the coil is represented by charges $- |q|$ with a speed of $v_q$. Their
influence on the passing electron is approximated by averaging over the angle
$\phi$ to find the mean $y$~component $\langle{v_{q,y}}\rangle$ of
$\mbox{\boldmath$v$}_q$ and an effective value for $b$. CG$_1$ indicates the
``centre of gravity'' at an effective impact parameter~$b^-_{\rm eff}$ for
the semicircle $\phi = 0~{\rm to}~\pi$ weighted with the $x$~component of
$v_q$. CG$_2$ at $b^+_{\rm eff}$ is the corresponding centre for
$\phi = \pi~{\rm to}~2\,\pi$.
\label{cross_section}}
\end{figure}
%
%
\begin{equation}
I = - \frac{Z\,|q|\,v_q}{2\,\pi\,R} ~ .
\label{current}
\end{equation}
The positive charges in the conductor are at rest. With $I = 0$ ($v_q = 0$),
an electron passing on the outside of the coil will, of course, experience no
forces. With $I \ne 0$ and thus $v_q \ne 0$, the situation changes by
relativistic time dilation in such a way that the density of the negative
charges $- |q|$ is dependent on their relative velocities of with respect to
the electron \citep{Dwietal}. We first treat one turn of the solenoid in the
$x$-$y$-plane, see Fig.~\ref{cross_section}. The $y$~component of $v_q$ is
$v_{q,y} = v_q\,\sin\phi$; integration over a half-circle with radius $R$
gives
%
\begin{equation}
M = R\,v_q\,\int^{\pi}_{0} \sin\phi~{\rm d}\phi =
2\,R\,\,v_q ~ ,
\label{mass}
\end{equation}
and the $x$~coordinates of CG$_1$ and CG$_2$ (cf., Fig.~\ref{cross_section})
are
%
\begin{equation}
\xi^\pm =
\pm\,\frac{R^2\,v_q}{M}\,\int^{\pi}_{0}\sin^2\phi~{\rm d}\phi =
\pm\,\frac{\pi\,R}{4} ~ ,
\label{cg}
\end{equation}
respectively. We define
%
\begin{equation}
b^{\mp}_{\rm eff} = b - \xi^\pm = b \mp\,\frac{\pi\,R}{4}
\label{b_eff}
\end{equation}
as effective impact parameters. Next we calculate a mean value of the
$y$~component of $v_q$
%
\begin{equation}
\langle{v_{q,y}}\rangle =
\frac{v_q}{\pi}\int^{\pi}_{0}\sin\phi~{\rm d}\phi =
\frac{2\,v_q}{\pi}
\label{v_mean}
\end{equation}
over $\phi$ from 0 to $\pi$, and
$\langle v_{q,y} \rangle = - 2\,v_q/\pi$ over $\pi$ to $2\,\pi$.
The corresponding mean $x$~components are zero. The momentum transfer between
charges $Q_1$ and $Q_2$ in relative motion is according to Equation~(10) of
\citet{Dwietal}
%
\begin{equation}
\Del P_b =
\frac{Q_1\,Q_2}{2\,\pi\,\varepsilon_0\,b\,v}
\label{Ppair}
\end{equation}
with $\varepsilon_0$ the electric constant in vacuum.

We assign portions of the negative and positive charges in the conductor to
the origins CG$_1$ and CG$_2$ of the effective impact parameters, considering
that the different speeds, $v_{\rm e} \pm 2\,v_q/\pi$, of the negative charges
with respect to the electron passing at $x = b$ must lead to an uneven
partition of the positive and negative charges at these positions. Evaluating
Eq.~(\ref{Ppair}) separately for $\phi$ from 0 to $\pi$ and $\pi$ to $2\,\pi$
provides approximations of the charge distributions in the winding close to
the $(x,y)$ plane as seen from the electrons on either side of the solenoid.
This is done in the following equation by taking the relativistic time
dilations \citep{Ein05} cause by the motions into account before the
determination of the momentum transfer to the electrons in the $x$ direction.
The speed $v$ has to be taken as $v_{\rm e}$ in this average configuration:
%
\begin{eqnarray}
\Del P_b^\mp (0)  \approx
\mp\,\frac{|e|}
{2\,\pi\,\varepsilon_0\,b^\mp_{\rm eff}\,v_{\rm e}}\,\frac{Z\,|q|}
{2}\,(\gamma - \gamma^\mp) ~ ,
\label{DPp}
\end{eqnarray}
where the upper indices of $\Del P_b^{\mp}$
refer to the influence of the charges $\Del Q^\mp_{\rm eff}$ defined by
%
\begin{eqnarray}
\Del Q^\mp_{\rm eff} = \frac{Z\,|q|}
{2}\,(\gamma - \gamma^\mp)
\label{charges}
\end{eqnarray}
at CG$_1$ and CG$_2$, respectively, with the notations for the Lorentz
factor related to the positive charges
%
\begin{eqnarray}
\gamma =
\left(1 - \frac{v^2_{\rm e}}
{c^2_0}\right)^{-1/2} \approx
1 + \frac{1}{2}\,\frac{v^2_{\rm e}}{c^2_0}
\label{Lorentz1}
\end{eqnarray}
and a factor of
%
\begin{eqnarray}
\gamma^\mp =
\left[1 - \frac{(v_{\rm e} \mp\,|\langle v_{q,y} \rangle|)^2}
{c^2_0}\right]^{-1/2} \approx
\nonumber \\
1 + \frac{1}{2}\,\left(\frac{v_{\rm e} \mp\,|\langle v_{q,y} \rangle|}
{c_0}\right)^2  ~ ,
\label{Lorentz2}
\end{eqnarray}
for the negative charges. $c_0$ is the speed of light in vacuum.
An evaluation with $v_{\rm e} \gg v_q$ yields
%
\begin{equation}
\Del Q^\mp_{\rm eff} =
\pm\,\frac{Z\,|q|\,v_{\rm e}\,v_q}{\pi\,c^2_0} ~ .
\label{approxcharge}
\end{equation}
If ${\rm d}N = 1$ represents one turn of the solenoid shown in
Fig.~\ref{configuration} the total effective charge, $Q_{\rm eff}$, can be
approximated by an integration along the length $a$ in the $z$-direction
%
\begin{eqnarray}
Q^\mp_{\rm eff} \approx
\int_{-\infty}^\infty \Del Q^\mp_{\rm eff}\,\frac{b}
{r}\,\cos\psi~{\rm d}N =
\nonumber \\
\Del Q^\mp_{\rm eff}\,\int_{-\infty}^\infty \cos^2\psi~{\rm d}N =
\nonumber \\
n\,b\,\Del Q^\mp_{\rm eff}\,\int_{-\pi/2}^{\pi/2}\,{\rm d}\psi =
\pi\,n\,b\,\Del Q^\mp_{\rm eff} ~ ,
\label{charge_int}
\end{eqnarray}
where we have used
%
\begin{eqnarray}
a = b\,\tan\psi = \frac{N}{n}
\label{tan}
\end{eqnarray}
and, after differentiation,
%
\begin{eqnarray}
{\rm d}N = \frac{n\,b}{\cos^2 \psi}\,{\rm d}\psi ~ .
\label{d_N}
\end{eqnarray}
Eqs.~(\ref{DPp}) and (\ref{charges}) together with Eqs.~(\ref{approxcharge})
and (\ref{charge_int}) yield after a short calculation
%
\begin{eqnarray}
P_b^\mp \approx
\mp\,\frac{\mu_0\,|e|\,n\,b\,Z\,|q|\,v_q}{2\,\pi\,b^\mp_{\rm eff}} ~ .
\label{DP}
\end{eqnarray}
With $\pi\,R^2 = S$, the area of the solenoid cross-section;
the magnetic flux density, $B_{\rm i} = \mu_0\,n\,I$;
as well as Eqs.~(\ref{current})
and (\ref{b_eff}), the total momentum transfer is:
%
\begin{equation}
P^*_b = P_b^- + P_b^+ \approx
-\frac{e\,S\,B_{\rm i}}{2\,b\,[1 -
(\pi\,R)^2/(4\,b)^2]} ~ .
\label{P_res}
\end{equation}

A corresponding calculation for an electron passing the solenoid at $x = - b$
gives the same result. The axial components cancel each other for reasons of
symmetry, see Fig.~\ref{configuration} and Eq.~(\ref{charge_int}).
It might be in order to mention that the slightly asymmetric configuration
during the approach leads to a minute differential momentum transfer according
to Eq.~(16) of \citet{Dwietal}. It can, however, be neglected against the
momentum of an electron of typically 30~keV. This can be confirmed by
considering the relative variation of the de Broglie wavelength during the
flyby due to the change of the refractive index \citep[cf.,][]{Gla33,EhrSid}.
It is of the order of $\approx 10^{-28}$.
%
\begin{figure}[t]
\vspace*{2mm}
\begin{center}
\includegraphics[width=8.3cm]{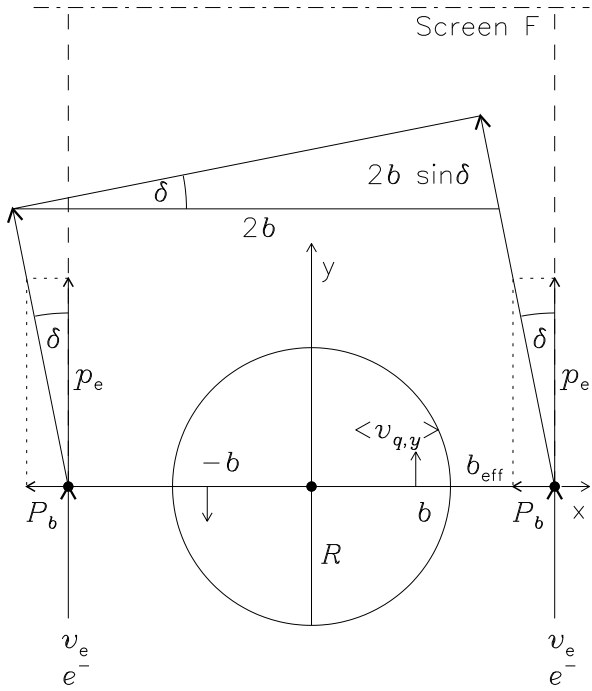}
\end{center}
\caption{Geometry of the electron diffraction near a solenoid.
Electrons with momentum $p_{\rm e}$ pass at impact parameters $\pm\,b$ with
respect to the origin of the $(x,y)$~coordinate system.
The momentum changes are $P_b$, each.
The deflection angle $\delta$ on either side
leads to a shift of the interference
pattern on the screen F\,---\,an appropriate focussing device is assumed,
for instance, an electro-optic
bi-prisma \citep{MoeDue}.
\label{solenoid}}
\end{figure}

Electrons with a momentum
%
\begin{equation}
p_{\rm e} = \frac{h}{\lambda_{\rm e}} ~ ,
\label{p_p}
\end{equation}
where $\lambda_{\rm e}$ is the de Broglie wavelength, passing
in Fig.~\ref{solenoid} at $x = \pm\,b$ will thus be diffracted according to
%
\begin{equation}
\tan \delta = - \frac{P^*_b}{p_{\rm e}} ~ .
\label{delta}
\end{equation}
The fringes of the interference pattern will be shifted by one order
on the screen~F, if
%
\begin{equation}
2\,b\,\sin \delta_1 = \lambda_{\rm e} ~ ,
\label{delta_0}
\end{equation}
or\,--\,with $\sin\delta_1 \approx \tan\delta_1$ for the small
angle $\delta_1$ under consideration and
Eqs.~(\ref{P_res}) to (\ref{delta_0})\,--\,if
%
\begin{equation}
S~[B_{\rm i}]_1 \approx  \frac{h}{e}\,[1 - (\pi\,R)^2/(4\,b)^2] ~ .
\label{AB1}
\end{equation}
For $b/R \gg 1$ it is
%
\begin{equation}
S~[B_{\rm i}]_1 \approx \frac{h}{e} ~ ,
\label{AB2}
\end{equation}
where $[B_{\rm i}]_1$ denotes the magnetic field required for a shift of
one order.

\section{Conclusion} 
\label{concl}

The electric dipole model \citep{Wiletal} and its application to magnetostatic
configurations \citep{Dwietal} provide an explanation of the Aharonov--Bohm
effect based on relativistic time dilations with reasonable accuracy
considering that\,--\,despite the many approximations made\,--\,the result in
Eq.~(\ref{AB1}) is for $b > 2\,R$ within a factor of $\approx 1.2$ of the
expected value given in the literature \citep[e.g.][]{AhaBoh,Capetal}.

\end{document}